\documentclass[compsoc, conference, a4paper, 10pt, times]{IEEEtran}

\usepackage{cite}
\usepackage{amsmath,amssymb,amsfonts}
\usepackage{algorithmic}
\usepackage{graphicx}
\usepackage{textcomp}
\usepackage{xcolor}
\usepackage{booktabs}
\usepackage[hidelinks]{hyperref}

\usepackage{algorithm}
\usepackage{array}
\usepackage{subfigure}

\usepackage{textcomp}
\usepackage{url}
\usepackage{verbatim}
\usepackage{soul}

\usepackage{comment}
\usepackage{balance}
\usepackage{breakurl}
\usepackage{makecell}
\usepackage{multirow}

\newcommand{\parag}[1]{\noindent\textbf{#1. }}

\begin{document}

\title{Hyperloop: A Cybersecurity Perspective}

\author{\\\IEEEauthorblockN{Alessandro Brighente}
\IEEEauthorblockA{\textit{University of Padova} \\
Padua, Italy \\
alessandro.brighente@unipd.it}
\and
\IEEEauthorblockN{Mauro Conti}
\IEEEauthorblockA{\textit{University of Padova} \\
Padua, Italy \\
TU Delft\\
Delft, The Netherlands\\
mauro.conti@unipd.it}
\and
\\\IEEEauthorblockN{Denis Donadel}
\IEEEauthorblockA{\textit{University of Padova} \\
Padua, Italy \\
denis.donadel@phd.unipd.it}
\and
\IEEEauthorblockN{Federico Turrin}
\IEEEauthorblockA{\textit{University of Padova} \\
Padua, Italy \\ 
SpritzMatter srl\\
Padua, Italy\\
federico.turrin@spritzmatter.com}
}

\maketitle

\begin{abstract}
Hyperloop is among the most prominent future transportation systems. It involves novel technologies to allow traveling at a maximum speed of 1220km/h while guaranteeing sustainability. Due to the system's performance requirements and the critical infrastructure it represents, its safety and security must be carefully considered. In transportation systems, cyberattacks could lead to safety issues with catastrophic consequences for the population and the surrounding environment. 
To this day, no research investigated the cybersecurity issues of the Hyperloop technology.

In this paper, we provide the first analysis of the cybersecurity challenges of the interconnections between the different components of the Hyperloop ecosystem. We base our analysis on the currently available Hyperloop implementations, distilling those features that will likely be present in its final design. 
Moreover, we investigate possible infrastructure management approaches and their security concerns. 
Finally, we discuss countermeasures and future directions for the security of the Hyperloop design.

\end{abstract}


\section{Introduction}

\begin{figure}[b]
 \centering
 \includegraphics[width=\columnwidth]{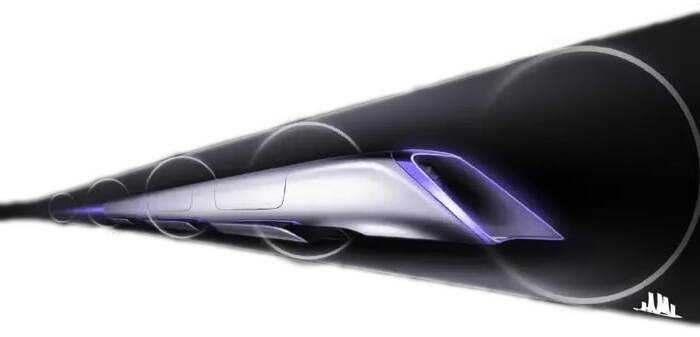}
 \caption{First Hyperloop conceptual design rendering~\cite{musk2013hyperloop}.}
 \label{fig:concept}
\end{figure}

Hyperloop is a radically new transportation system concept introduced in 2013~\cite{musk2013hyperloop} as a preliminary version of the Hyperloop project connecting Los Angeles to San Francisco. The initial vision, represented in Figure~\ref{fig:concept}, is to provide fast transportation means by mitigating air resistance and friction. 
Modern fast train technologies for mass transportation include High-Speed Trains (HSTs)~\cite{polunsky2017homeland} and Maglev~\cite{janic2021estimation}. HST includes trains that mechanically move on the railway and can reach up to $250$ km/h, while Maglev is a novel train generation that leverages magnetic levitation to reduce friction and achieve a velocity of up to $600$ km/h.
Hyperloop represents a radical new technology aiming to drastically increase transport speed by leveraging the absence of air resistance in vacuum environments. However, Hyperloop introduces novel design challenges~\cite{MTST81420}.
Besides the speed objective, Hyperloop also allows for reduced operating costs and, hence, for energy savings. After the first official document, the project continued through university contests organized by SpaceX and later with the foundation of startups. Today, companies are working on Hyperloop systems, and testbeds are deployed around the world~\cite{hyperloopWikipedia, hyperloopitalia, delft, htt}.

Hyperloop is characterized by two main elements: a tube with vacuum capabilities and a capsule that travels along the tube with passengers or cargo, exploiting air pressure and electromagnetic forces. Hyperloop promises a set of advantages compared to available public transportation systems. First, it provides high-speed transportation. Airlines generally fly at $980$km/h, and special trains reach a speed of $600$km/h, while the expected for Hyperloop is $1220$km/h. Second, it allows for reduced operational costs. By placing solar panels over the tube carrying the capsule, the generated energy can be employed to run the system~\cite{musk2013hyperloop}.

In recent years, cybercriminals are increasingly targeting cyber-physical systems~\cite{duo2022survey, thakur2016impact}. These include attacks against nuclear plants~\cite{falliere2011w32} and vehicles and transportation systems~\cite{reutershacker}. Therefore, assessing the system's weaknesses is necessary to prevent dangerous consequences. The main distinguishing feature of Hyperloop compared to other transportation means is its increased automation, achieved by establishing communication links among its components. 
System entities such as pods and infrastructure exchange critical information the Hyperloop ecosystem uses to guarantee reconfigurability and safety. 
To this day, no study on Hyperloop focused on the security and privacy implications of its communications. It represents a fundamental need to be met before its deployment.

\parag{Contributions}
This paper analyzes the current state-of-the-art Hyperloop technology components and the cybersecurity challenges arising from their interactions. Despite the lack of official documentation, we collected information from publicly available sources and identified common elements with other transportation systems, such as aviation, railways, and automobiles, to create a high-level description of the Hyperloop system. Based on similar domain knowledge, we focus on the different components of the Hyperloop and their interconnection, identifying the critical security and privacy vulnerabilities they imply. By considering the overall Hyperloop ecosystem as a group of interconnected sub-systems, we focus on their interaction via signaling and communication. 
This provides us with means to analyze the cyber-physical security of Hyperloop at different levels, from in-pod threats to complex vehicle-to-infrastructure interactions. 
Finally, we propose security practices to prevent possible attacks. To the best of our knowledge, this is the first work focusing on the cybersecurity challenges related to Hyperloop technology. 
Notice that we decided to focus on the interaction of Hyperloop's components instead of focusing on the cybersecurity of its components. For instance, we did not examine possible attacks on Hyperloop's controllers. Despite this, our analysis provides a first step toward future studies in developing a secure Hyperloop ecosystem.

\parag{Organization} We start this work by presenting some related works in Section~\ref{sec:related} and the overall Hyperloop infrastructure in Section~\ref{sec:infrastructure}. Then, in Section~\ref{sec:network}, we analyze the communication between the various components of the system deeply. Section~\ref{sec:analysis} investigates the security issues of the Hyperloop system, while section~\ref{sec:countermeasures} propose effective countermeasures to the discovered attacks. Finally, Section~\ref{sec:conclusion} concludes our work with some final insights.

\section{Related Work}\label{sec:related}

The Hyperloop, a cutting-edge technology in its nascent stages, currently lacks an extensive body of literature and documentation. Nevertheless, several studies actively explore the unique challenges posed by Hyperloop technology~\cite{noland2021prospects, chafii2023twelve}. Elon Musk's seminal document from 2013 serves as the foundational source, providing a comprehensive vision of the system~\cite{musk2013hyperloop}. Gkoumas~\cite{gkoumas2021hyperloop} proposed a survey on the academic research covering different aspects of the Hyperloop system, from communication to aerodynamic and energy issues. As shown, while various papers discussed safety issues on Hyperloop~\cite{thakur2016impact, brusyanin2014basic}, security is never investigated. 

\parag{Testbed}
While Hyperloop technology is still in its early stages, numerous testbeds have been deployed worldwide to conduct preliminary experiments. Notably, Hyperloop TT~\cite{htt} and Virgin Hyperloop~\cite{virgin} stand out, with the former conducting the world's first passenger journey in 2021. Testbeds have gained momentum globally, reaching an advanced level of maturity in countries such as The Netherlands~\cite{delft}, South Korea~\cite{korea}, Italy~\cite{hyperloopitalia}, Spain~\cite{spain}, Poland~\cite{poland}, France~\cite{france}, and Switzerland~\cite{swiss}. However, testbeds are focused on testing the main challenges of the system, such as the physical structure and mechanical components, partially leaving aside communication aspects.

\parag{Infrastructure Challenge}
A multitude of research focuses on the structural and physical challenges involved in developing Hyperloop infrastructure~\cite{noland2021prospects, gkoumas2021evidence}. The extreme operating conditions of Hyperloop, such as high speeds and vacuum, necessitate cutting-edge solutions to design a transportation system that is both reliable and safe. Researchers analyze various physical aspects of the infrastructure, including aerodynamic structures~\cite{opgenoord2018aerodynamic, nick2020computational, nowacki2019assessment}, levitation~\cite{seo2020study, chaidez2019levitation},  thermal flows~\cite{heat}, sustainability~\cite{riviera2018high}, and safety~\cite{van2018analysis, mateu2021setting}.

\parag{Network Challenges}
Instead, only a few studies focus on the communication infrastructure of Hyperloop, even if several challenges need to be addressed. For instance, ultra-high-speed brings challenges related to wireless communication because of the frequent handovers and to the management of a serious Doppler frequency shift~\cite{han2020wireless, qiu2020broadband, zhang2020concepts}. Tavsanoglu \textit{et al.} delve into the internal communication system, identifying 802.11ax networks and 5G NR as the current sole viable communication technologies for Hyperloop speeds when considering already mature technologies~\cite{tavsanoglu2021concepts}. However, other solutions have been proposed and developed in the last few years. For instance, Huang \textit{et al.} proposed a handover-free method employing optical wireless communication~\cite{huang2021optical}. Qiu \textit{et al.} propose a comprehensive analysis of the design challenges of such a communication system and propose an architecture based on a cloud Radio Access Network (C-RAN)~\cite{qiu2020broadband}.
Other details on the communication have been investigated as well, such as the employment and positioning of intelligent reflecting surfaces~\cite{hedhly2023intelligent}, or the scheduling of mechanisms when using millimeter wave communications~\cite{li2024scheduling}. Moreover, Hedhly \textit{et al.} focus on designing the Hyperloop network architecture, addressing quality-of-service requirements and the associated challenges in implementing a reliable communication system~\cite{hedhly2021hyperloop}. 

While some other papers briefly mentioned a physical security issue~\cite{dudnikov2017advantages, MTST81420}, our focus centers on the cybersecurity aspects of Hyperloop technology, drawing parallels with systems of a similar class, such as CPSs and other high-speed trains. To the best of our knowledge, this study is the first to explore the cybersecurity landscape of Hyperloop.

\section{Overview of the Hyperloop Infrastructure}\label{sec:infrastructure}

In this section, we overview the components of the Hyperloop system, focusing first on the physical infrastructure (Section~\ref{subsec:phyinfra}) and then on the network infrastructure (Section~\ref{subsec:netinfra}).

\subsection{Physical Infrastructure}\label{subsec:phyinfra}

\begin{figure}[t]
    \centering
  \begin{subfigure}[The Tube]
  {\includegraphics[width=\columnwidth]{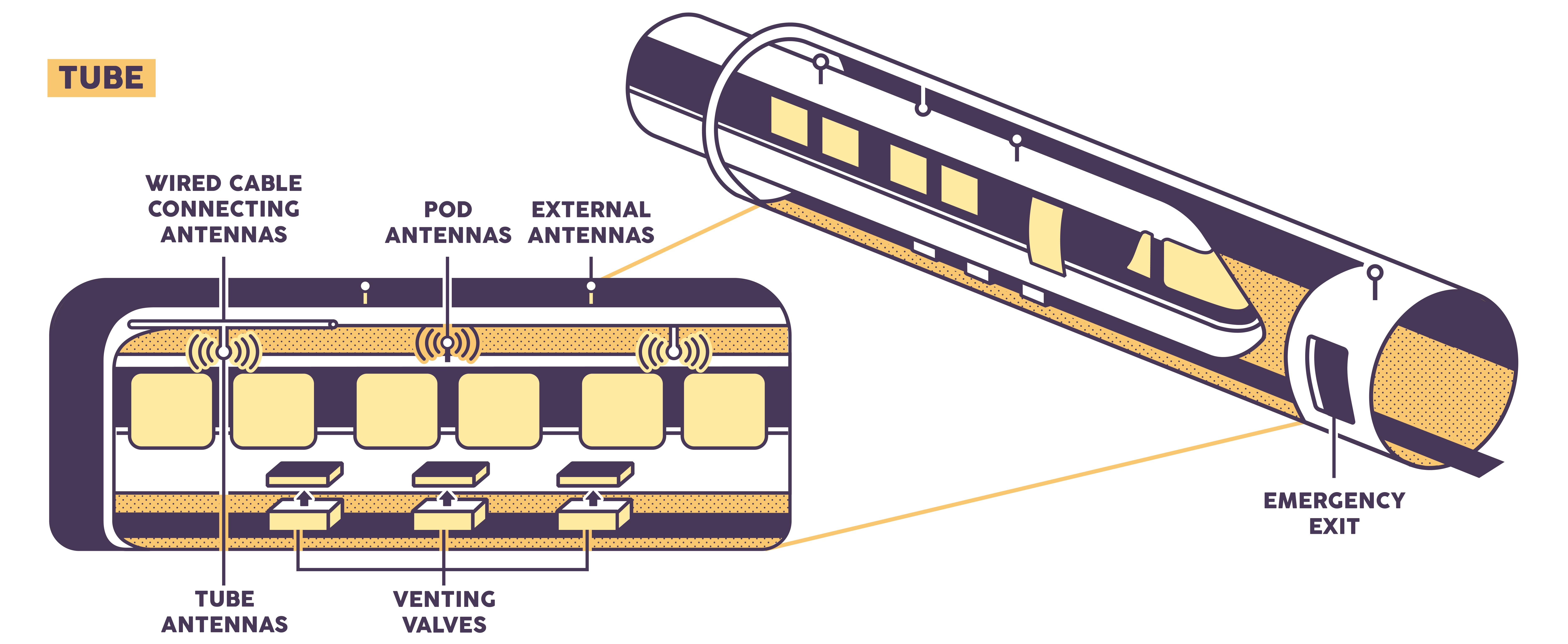}\label{subfig:tube}}
  \end{subfigure}
    \\
  \begin{subfigure}[The Capsule]
  {\includegraphics[width=\columnwidth]{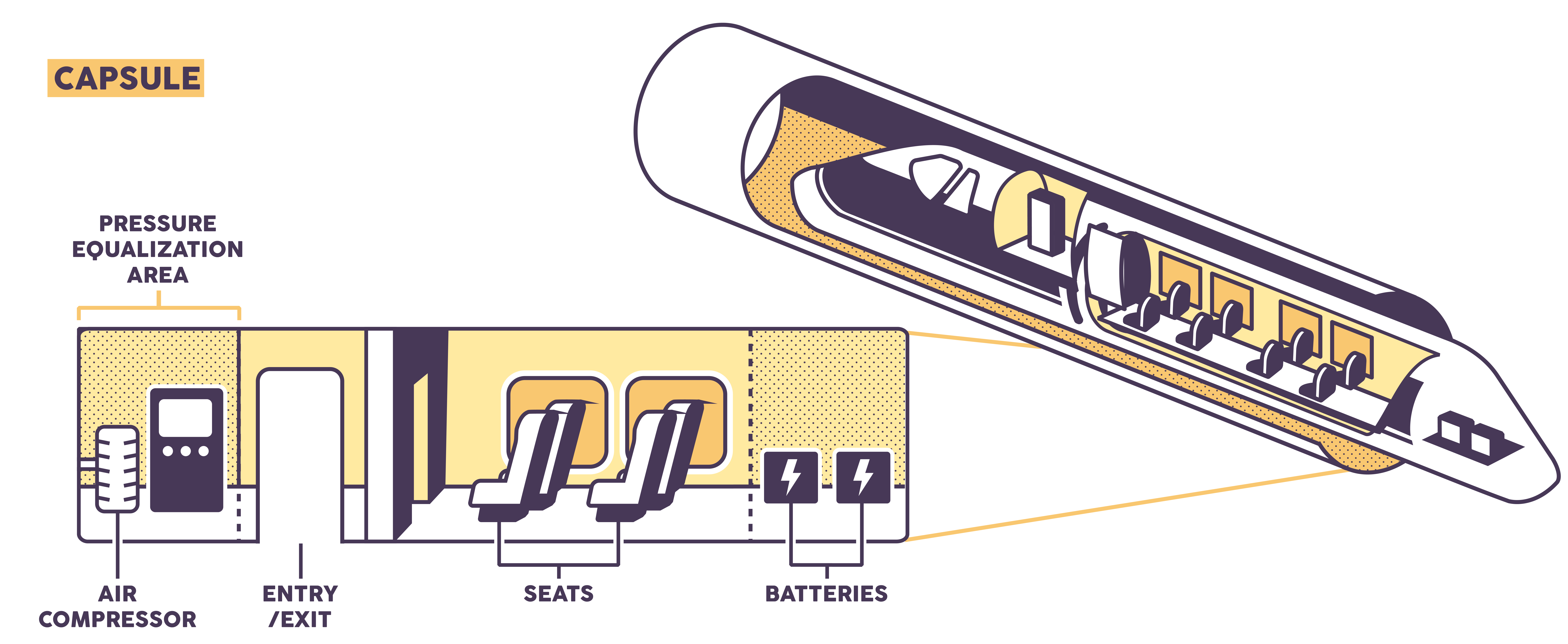}\label{subfig:capsule}}
  \end{subfigure}
  \\
  \begin{subfigure}[The Station]
  {\includegraphics[width=\columnwidth]{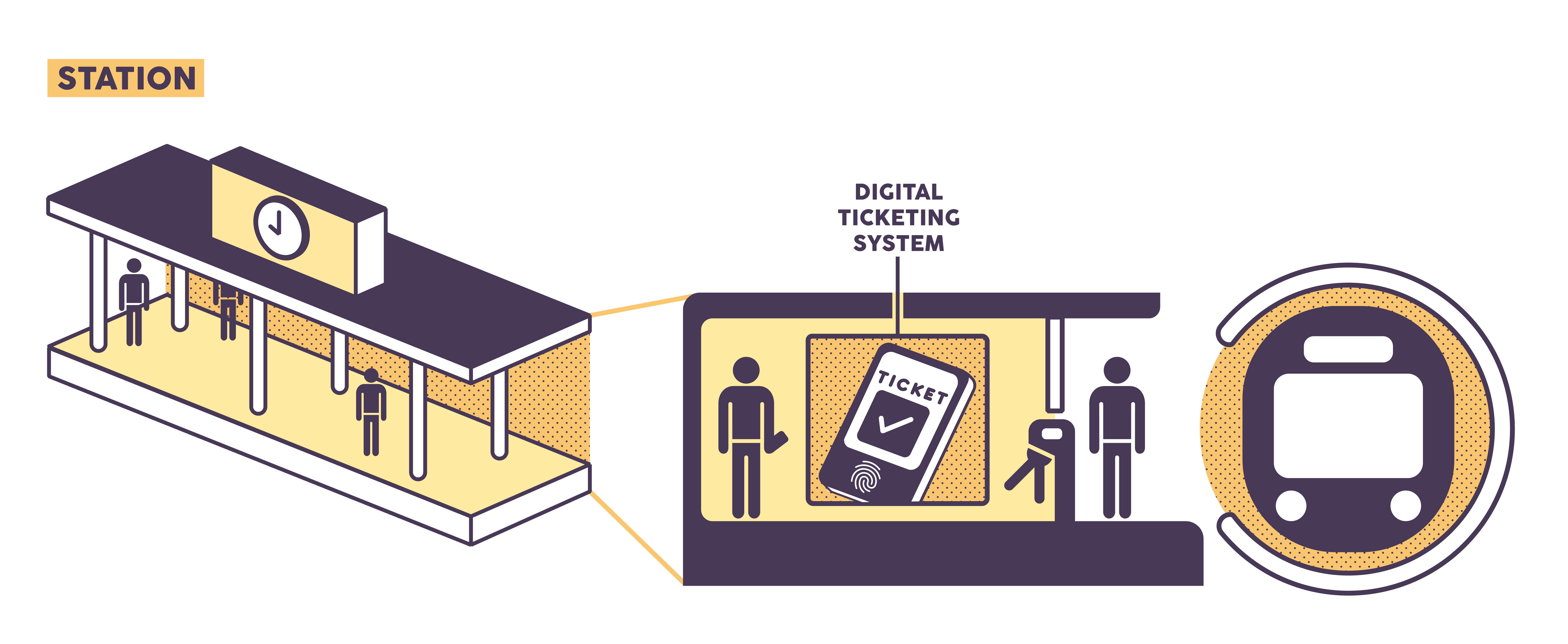}\label{subfig:station}}
  \end{subfigure}
   
  \caption{Overview of the Hyperloop physical infrastructure.}
  \label{fig:infrastructure} 
\end{figure}

The physical infrastructure of the Hyperloop system comprises three macro components: the tube, the station, and the capsule. Figure~\ref{fig:infrastructure} details the components and the various devices included in each of them. 

\parag{Tube} The tube is a pipe creating a near-vacuum environment reaching an air pressure equivalent to $60,960$m above sea level, where the Hyperloop capsule transits with reduced aerodynamic drag. The details are depicted in Figure~\ref{subfig:tube}.
The tube is equipped with emergency egresses every $75$m to allow passengers to exit the vehicle in case of need~\cite{virgin}. The tube uses valves to isolate and re-pressurize part of the tube to ease the repairing process.
The Hyperloop system aims at producing more energy than it consumes thanks to solar panels placed over the tube~\cite{musk2013hyperloop, virgin}. The energy consumption of the Hyperloop mainly depends on the journey's distance and the number of seats, but it is generally lower than aircrafts~\cite{janic2021estimation}. 

\parag{Capsule} Hyperloop trains are called pods or capsules. Figure~\ref{subfig:capsule} schematizes the pod’s structure and components.
The capsule is the size of a small commercial aircraft with a capacity of $28$-$50$ passengers each~\cite{chin2015open}. Different levitation mechanisms are under investigation, ranging from air beaming employed in the first concept~\cite{musk2013hyperloop} to more efficient electromagnetic or electrodynamic suspension systems~\cite{noland2021prospects}.
Capsules are designed for ultra-high-speed using various technologies. In particular, pods can be wrapped with materials to monitor and transmit critical information such as temperature, stability, and integrity. Two or more lithium-ion battery packs power the capsule life support system to prevent power outages.

\parag{Station} According to~\cite{virgin}, the Hyperloop portal will be a central hub that integrates all nearby transportation systems to enable a seamless end-to-end journey. The station will offer digital ticketing, biometric check-in, wayfinding, and on-demand boarding services. The Hyperloop station will include portals from which passengers may depart and arrive. Figure~\ref{subfig:station} shows a simplified representation of the station.

\subsection{Network Infrastructure}\label{subsec:netinfra}
\begin{figure}[bt]
 \centering
 \includegraphics[width=0.9\columnwidth]{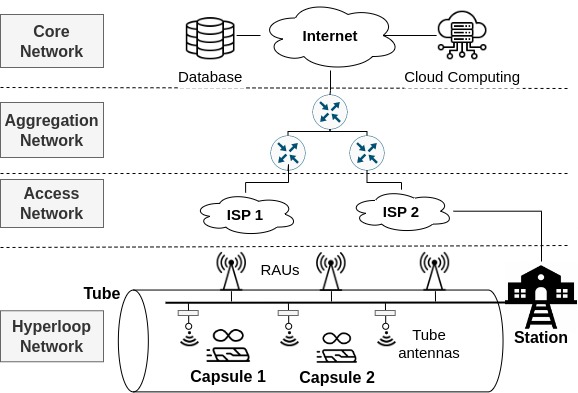}
 \caption{Network layers composing the Hyperloop Network Infrastructure.}
 \label{fig:system}
\end{figure}

The network architecture conceptualized to manage Hyperloop is proposed in different works~\cite{tavsanoglu2021concepts, hedhly2021hyperloop}. As depicted in Figure~\ref{fig:system}, we divide the network infrastructure into four layers.

\parag{Hyperloop Network} This Hyperloop-specific layer contains the first communication link with the field network. This includes communications between capsules (originating from the pod itself or user services), between capsules and the station, or between capsules and the Internet. To support the fast communication between internal antennas, Remote Access Units (RAUs), and stations, we can assume the existence of a wired bus inside the tube that interconnects these entities. RAUs are placed outside the tube to provide an Internet connection to the communication inside the tube. The connection between the Hyperloop Network and the Access Network can be (e.g., WiFi, 5G) or wired (e.g., optical fibers) to meet the requirement of fast communication. The tube is internally equipped with several antennas to communicate with the pods. More antennas are installed on capsules to alleviate the timely expensive handover process~\cite{tavsanoglu2021concepts}. The communications originated by the users inside the pods and from the instrumentation on the capsule are captured by the internal antennas and then forwarded to the station (via a wired connection) or the Internet (via RAUs).

\parag{Access Network} Connects the Hyperloop Network and the Aggregation Network. The Access Layer is also responsible for gathering the information between the various RAUs placed along the tube and connecting them with one or more Internet Service Providers (ISPs). In fact, since the tube can be hundreds of kilometers long through different countries, the ISPs can differ based on the geographical area. An ISP is an organization that provides the Internet connection to the requestor by assigning it a public IP address.
Besides communication coming from the pods, this network allows the ground station of Hyperloop to connect to the Internet, enabling services such as remote connection

\parag{Aggregation Network} It connects the Access Network to the Core Network by gathering and routing the information between the different ISPs. ADSL, WiFi, Ethernet, and optical fiber are among the technologies used in this layer.

\parag{Core Network} The purpose of this layer is to interconnect the different components of the network, enabling access to the various facilities and services, such as databases and clouds, for computations. It must provide low congestion, short delays, high availability, and strong adaptability to support future applications. 

\section{Hyperloop Communication Infrastructure}\label{sec:network}

\begin{figure}[tb]
 \centering
 \includegraphics[width=.9\columnwidth]{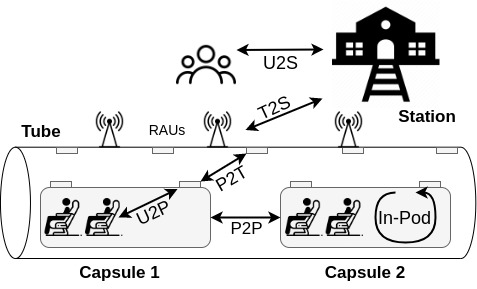}
 \caption{Communications types inside the Hyperloop Network Layer.}
 \label{fig:network}
\end{figure}

In the following, we divide the network into the communication channels employed for the system’s operation. Figure~\ref{fig:network} represents a schema of the various communications in the Hyperloop network.
We divide communications into Pod-to-Pod (P2P) (Section~\ref{subsec:p2p}), Pod-to-Tube (P2T) (Section~\ref{subsec:p2t}), User-to-Pod (U2P) (Section~\ref{subsec:u2p}), Tube-to-Station (T2S) (Section~\ref{subsec:t2s}), In-Pod (InP) (Section~\ref{subsec:inp}), and User to all devices available in the stations (U2S), such as the ticketing service, discussed in Section~\ref{subsec:u2s}.

\subsection{Pod-to-Pod Communication}\label{subsec:p2p}

P2P communications are similar to vehicle platooning~\cite{ghosal2021truck}. Pods are expected to automatically merge and detach based on their routes. To avoid collisions, a pod must detect the presence of another pod and its relative location. To this aim, pods broadcast messages containing information on their location and speed. Data transmission between high-speed pods could be supported by ad-hoc protocols similar to Dedicated Short Range Communications used for vehicle-to-vehicle communication.

\subsection{Pod-to-Tube Communication}\label{subsec:p2t}

The P2T link is used to control pods. The tube can control and manage the magnetic or air forces needed to handle the pod's movements and maneuvers through this link. Vice versa, the pod can apply magnetic or air forces for the movement generated by the onboard engine. %
The P2T link can also be used to exchange messages about the status of the capsule and its location. This information will then be delivered via a T2S link to the central station for managing purposes, such as pressure regulation and motor management. Thanks to RAUs deployed along the tube, the pod can use the P2T link to access the Internet to guarantee users' connectivity or exchange information with external entities~\cite{hedhly2021hyperloop}.

As discussed in~\cite{tavsanoglu2021concepts}, only two wireless standards support communication at the Hyperloop speed: 802.11ax networks and 5G NR. They are, therefore, likely to be employed in the Hyperloop infrastructure between internal antennas and pods.

\subsection{User-To-Pod Communication}\label{subsec:u2p}

The pod is equipped with an internal wireless network to provide an infotainment system to users during transportation. Hence, users can retrieve information regarding the trip status, their current location, and expected arrival time. Furthermore, users can access entertainment material and, more generally, the Internet. This connection is guaranteed by the P2T communication link that provides access to the Internet, thanks to the RAUs located along the tube. The access points are connected to the local ISP, which enables the connection to the core network providing Internet access.

\subsection{Tube-to-Station Communication}\label{subsec:t2s}

The operations that the tube needs to handle to guarantee the pods' safety during travel must be managed by a central entity. These operations include operating valves that create the near-vacuum environment and opening emergency exits in case of need. Furthermore, the tube needs to actuate the pods' motion according to the scheduling and report information about the pods' status to the station for managing purposes. These connections can be wired or wireless, depending on the requirements in terms of latency and reliability. Furthermore, a dedicated power line will deliver the energy collected via solar panels to the accumulator deployed at the central station. A second power line will provide energy for operations related to the tube. These operations include the application of power for levitation and pod direction.

\subsection{In-Pod Communication}\label{subsec:inp}

Each pod has an internal network where controllers and actuators regulate the basic pod functioning. This concept is similarly applied  in modern vehicles, where a Controller Area Network bus connects many Electronic Controllable Units (ECUs). The pod might use a similar technology to e.g., manage air conditioning, door operations, fire alarms, and battery management system~\cite{singh2021board}. All these units are connected to the pod safety controller that continuously monitors data. This network is also used to deliver power to the different components of the pod. For instance, suitable controllers handle the flank propulsors that allow for high-speed movements. 

\subsection{User-to-(Devices in the) Station Communications}\label{subsec:u2s}

For a complete analysis of the Hyperloop system, we must consider the communications occurring inside the station between the users and the infrastructure. Tickets can be bought from ad-hoc terminals, which have to verify the availability of the seats on the pods in real time. Turnstiles manage the user's access to the boat area and could use the user's biometrics information to facilitate access. Displays show delays and departure times for pods and help users find the proper gate. Furthermore, even if not physically located in the stations, we can consider the smartphone apps and web services that enable users to buy tickets remotely in advance to avoid long queues.

\section{Cybersecurity Analysis}\label{sec:analysis}

Based on similar domain knowledge (e.g., industrial and vehicular systems), we present an analysis of the possible security and privacy issues for each communication category. In particular, Section~\ref{subsec:a_p2p} focus on P2P security, Section~\ref{subsec:a_p2t} on P2T security, Section~\ref{subsec:a_u2p} on U2P security, Section~\ref{subsec:a_t2s} on T2S security, Section~\ref{subsec:a_inp} on InP security, and Section~\ref{subsec:a_u2s} to U2S security.
Moreover, Table~\ref{tab:secuirty} summarized the different attacks affecting each communication link, the corresponding threats, and their effect on the system.

\subsection{Pod-to-Pod Security}\label{subsec:a_p2p}

P2P communication is not available in HST or Maglev systems. Instead, Hyperloop offers the possibility of attaching/detaching pods and driving them independently~\cite{virgin}, sharing some similarities with Vehicle-to-Vehicle communication in platoons~\cite{ghosal2021truck}. Several cyberattacks can affect this communication: Man-in-the-Middle (MitM) attacks can modify or craft packets to include false information about the pod and generate a bad reaction in the neighboring pods. For instance, one pod under attack can claim to be close to the preceding pod and make it speed up. Furthermore, an attacker might capture P2P communication packets in a blackhole attack preventing packet forwarding. These attacks, especially if coordinated on a certain number of capsules, can be very destructive and create disorder in the network. MitM can also send malicious messages to disconnect a pod from a convoy. Another way to force the disconnection of a pod could be a flooding attack leading to a Denial of Service (DoS). It could prevent a capsule from receiving messages from other pods, inhibiting early reactions to safety problems. 

Even if a complete disconnection of the target pod is not possible due to the high capacity of the communication channel, a slight delay in the packet transmission could be dangerous due to the hard real-time constraints on location sharing to avoid incidents. Although this is a problem in vehicle platoons as well, it is more relevant in Hyperloop because of the high speed reached by the system.

Supposing pods periodically broadcast their position to alert all nearby pods, spoofing a false location could make all receivers take wrong actions, e.g., break or speed up/down even if not needed, leading to inefficiency (e.g., delays) or crashes.

\subsection{Pod-to-Tube Security}\label{subsec:a_p2t}

P2T communication happens similarly in HST and Maglev. This communication has several purposes ranging from critical aims (e.g., managing the levitation of the pod) to secondary scopes (e.g., forwarding the traffic from the infotainment system of the pod). However, unlike other transportation systems, Hyperloop's infrastructure is in charge of controlling the pressurization of the tube. A MitM attack on it can have different impacts, ranging from slowing down the trains' speed to creating safety issues for the passengers during loading/unloading if doors are unsafely opened. Hijacking a pod may also be possible. In fact, due to high-speed switching, an attacker may trigger a switch to detach a pod from a convoy and redirect it in the wrong direction.

In Hyperloop, location data is transmitted P2P. However, similarly to other rail systems, the pod location is also transmitted to the ground station through the tube, for central safety management.
An attacker can spoof this communication to send messages with false locations of the pods to generate inconsistencies in the system or to hide the presence of a pod.
Although the pod's location might help schedule purposes and estimate arrival times, real-time location sharing might represent a privacy threat, which must be considered during the design phase.

The high speed of the Hyperloop and the isolation from the outside world generate a novel context where traditional communication channels may not work properly. Therefore, the pod must forward the data through the tube to reach the ground station and the ISP. An attacker could disable or delay this exchange of messages by performing a flooding attack from a pod. DoS could lead to the disruption of the infotainment system and the delay of safety-critical data flows. Moreover, the P2T communication needs frequent handover phases, which, if not correctly configured, can be a profitable point for launching DoS attacks. Furthermore, because of the high speed, even short DoS or other attacks causing delays can significantly impact the system functionalities.

It is possible to imagine an exchange of energy between the tube and the pod (via, e.g., wireless power transfer) to charge the pod’s batteries. This unique energy flow must be managed by the tube following the requests coming from the pod. An attacker could tamper with this communication to stop the energy request and avoid charging the pod, or it may require more power than needed to damage the batteries. Furthermore, a DoS could prevent packet exchange, lead to unpredictable energy transfer behavior, or limit the number of pods to benefit from it.

\subsection{User-to-Pod Security}\label{subsec:a_u2p}

A central purpose of the U2P network is the infotainment system, similar to HST or Maglev. A malicious user can generate a DoS attack to prevent others from using the system. To access the infotainment system, the user is requested to authenticate using credentials linked to the ticket to assign bandwidth and fairly keep track of users' misconduct. A malicious actor can hijack another user to circumvent restrictions or DoS the authentication system to avoid other users connecting to the access point. However, since the pod is shielded inside the tube, stopping this communication channel means stopping all communication outside the tube.

Given the number of users connected to the same access point, a malicious entity could try to attack other users, for instance, to eavesdrop on communications and steal sensitive information. MitM could also be possible for non-encrypted communications, while traffic analysis could be applied to infer users' activities by monitoring encrypted data. With a view to the future, a router can also cache content to serve popular user requests faster~\cite{zhang2020concepts}. These systems can be vulnerable to cache poisoning attacks, leading to phishing attacks and credentials leakages. A malicious user can also eavesdrop on U2P communication to track and profile a specific user's pod access.
Moreover, the high interconnection of pods makes them more similar to airplanes than trains. This added to the fact that the connection with the outside world is possible only using the Hyperloop-provided service, makes it essential to create a highly segmented network to separate the infotainment connection from the safety channel. A malicious attacker can compromise the latter or slow down the connection via DoS.

\subsection{Tube-to-Station Security}\label{subsec:a_t2s}

Similarly to what happens with communication between HST and stations, T2S communication must be reliable since it has to manage the high traffic of pods and possible issues regarding trips. A DoS attack can undermine the system's availability and prevent the exchange of T2S messages.
However, because of the incredible speed reached by Hyperloop, even a tiny delay in data communication can be devastating in systems with hard real-time constraints if proper mitigation techniques are not in place.
Likewise, manipulating the transmitted data can damage the system, affecting the very specific Hyperloop pressurized tube. Sending malicious messages can affect its functioning, such as forcing depressurization when unnecessary. An evil entity who obtained control of the channel could also insert false anomalies or spoof pods’ information, for instance, to make the system stop for useless maintenance. On the contrary, if the system suffers from minor malfunctions, an attacker may spoof sensor values to hide anomalies and prevent the proper monitoring of the Hyperloop system. Another possible effect of an injection attack could be the complete shutdown of tube parts by, e.g., opening a closing depressurization valve, creating an interruption of the service and potential accidents.

Since the tube covers long distances, possibly in different states, it is also important to consider physical attacks which can damage the communications. A malicious actor can tamper with the wired connections to stop communication if cables are exposed to external entities. All the devices, such as RAUs placed along the tube, need constant monitoring because tampering with the T2S communication could lead to catastrophic consequences. 

\subsection{In-Pod Security}\label{subsec:a_inp}

Even if most of the management and computations occur outside the pod, such as in HST and Maglev, the capsule has to provide sensor measurements and manage the actuators to execute the station's orders. More importantly, the pod may require to manage its levitation.
A DoS in the in-pod network could lead to a consumption of resources and prevent or delay the reception of critical messages. These urgent messages control crucial systems such as emergency stops or fire detectors. Hence even a slight delay in the packets could lead to accidents. In addition to critical systems, as in HST and Maglev, the capsule controls non-critical devices such as ventilation and infotainment systems. Tampering with these technologies can create discomfort to the users (e.g., losses of connection, high/low air temperature) or compromise user privacy by eavesdropping on the communication between the users and the access point. An attacker gaining MitM capability in the in-pod network could propagate false messages, creating inconsistencies in the pod’s status and false alarm messages. Alternatively, the malicious user could lead to dangerous behaviors of the pod’s components (e.g., opening the door while the pod is in movement) or tampering with the battery management system to remove the power source.

\subsection{User-to-(Devices in the) Station Security}\label{subsec:a_u2s}

Hyperloop stations share the same concept as train system stations, thus inheriting most security issues. Thanks to MitM and eavesdropping, an evil user can steal sensitive data. Information on a specific user's habits may lead to profiling or user tracking, representing a privacy violation. An attacker may also spoof the U2S communication to create false messages. For instance, an attacker may be able to bill another user for a ticket or modify the information shared with a user on the availability of free seats in a pod, thus preventing purchases. 
Furthermore, an attacker may modify the information the station shares with the user on scheduling and pod tracking, thus causing traveling issues.

\begin{table*}[!ht]
\renewcommand{\arraystretch}{1.4}
\centering
\caption{Possible attacks and threats affecting Hyperloop and the consequent effect on the system.}
\label{tab:secuirty}
\resizebox{\textwidth}{!}{
\begin{tabular}{p{1.9cm}p{1.8cm}p{2.5cm}p{7cm}p{3cm}p{0.9cm}} \hline
\textbf{Communication} & \textbf{Attack} & \textbf{Threat} & \textbf{Effect} & \textbf{Countermeasure} & \textbf{Impact} \\ 
\hline
\multirow{3}{*}{\textbf{Pod-to-Pod}} & MitM, Spoofing, Relaying, Replaying & Impersonation, Information Gathering, Vulnerability Exploitation, Manipulation, System Corruption, Service Loss & \multirow{2}{*}{\makecell[{{p{7cm}}}]{-  Malicious message injection to make the adjacent pods break or speed-up/down to cause collisions \\ - Disconnect a connected pod, or connect a disconnected pod}} & Authentication, Encryption & High \\ 
\cline{2-6}
& DoS, Flooding & System Corruption, Service Loss & \multirow{2}{*}{\makecell[{{p{7cm}}}]{- Pod unable to receive messages from other pods\\- Generated delays make received information useless}} & Resource management recovery, IDS & Medium \\ 
\cline{2-6}
& Location Spoofing & Manipulation, System Corruption & - Inject false information on the location of nearby pods to create collisions & Distance bounding, Digital Signature & High \\ 
\hline
\multirow{4}{*}{\textbf{Pod-to-Tube}} & MitM, Relaying, Replaying & Impersonation, Information Gathering, Vulnerability Exploitation, Manipulation, System Corruption, Service Loss & \multirow{3}{*}{\makecell[{{p{7cm}}}]{- Sending fake commands to a pod or the tube\\- Derail a pod\\- Hijack a pod}} & Authentication, Encryption, IDS & High \\ 
\cline{2-6}
& DoS, Flooding & System Corruption, Service Loss & \makecell[{{p{7cm}}}]{- Disable or delay the exchange of messages between tube and pod \\- Prevent exchange of energy between the tube and the pod} & Resource management recovery, IDS & Medium \\ 
\cline{2-6}
& Location Spoofing & Manipulation, System Corruption & - Create inconsistencies between pod’s real and claimed locations & Distance bounding, Digital Signature & Medium \\ 
\cline{2-6}
& Privacy violation & Data Leakage & - Track the location of a pod, together with identifiers of onboard users & Data Anonymization, Encryption & Low \\ 
\hline
\multirow{3}{*}{\textbf{Tube-to-Station}} & MitM, Spoofing, Relaying, Replaying & Information Gathering, Vulnerability Exploitation, Manipulation, System Corruption, Service Loss & \multirow{4}{*}{\makecell[{{p{7cm}}}]{- Send malicious control messages that affect the tube’s functioning \\- Generate anomalies or spoofing the pod information\\- Shutdown parts of the tube\\- Hide running anomalies}} & Authentication, Encryption, IDS & High \\ 
\cline{2-6}
& DoS & System Corruption & - Prevent exchange of messages from T2S and vice-versa & Firewall, IDS & Medium \\ 
\cline{2-6}
& Physical Tampering & System Corruption, Service Loss & \multirow{2}{*}{\makecell[{{p{7cm}}}]{- Cut the wired communication\\- Tamper with the RAUs to interrupt the service}} & Enforcement with anti-tamper materials, Anomaly detector & Medium \\ 
\hline
\multirow{3}{*}{\textbf{User-to-Pod}} & MitM, Eavesdropping, Authentication system attack & Data Leakage, Service Loss & \multirow{3}{*}{\makecell[{{p{7cm}}}]{- Spoof users communications to the external network\\- Steal sensitive informations\\- Prevent pod's access with the ticket}} & IDS, Encryption, Authentication, Data Anonymization & Low \\ 
\cline{2-6}
& DoS, Authentication system attack & System Corruption, Service Loss & - Prevent user to exchanging information with the pod (mostly related to infotainment) and so with the Internet & Resource management recovery, Authentication, Firewall & Low \\ 
\cline{2-6}
& Privacy violation, Phishing & Data Leakage & - Steal private information of users in the pod & Encryption, Authentication & Low \\ 
\hline
\vspace{0.27cm}\multirow{2}{*}{\makecell[l]{\textbf{In-Pod-} \\ \textbf{Communication}}} & MitM, Spoofing, Relaying, Replaying & Information Gathering, Vulnerability Exploitation, Manipulation, System Corruption, Service Loss & - Compromise in-pod critical and non-critical systems (e.g., infotainment, light, fire detectors) & IDS, Integrity check & Medium \\ 
\cline{2-6}
& DoS, Flooding & System Corruption, Service Loss & - Consume resources and prevent or delay the reception of critical messages (e.g., malfunctions, emergency stops) & Firewall, IDS & High \\ 
\hline
\vspace{0.27cm}\multirow{3}{*}{\makecell[l]{\textbf{User-to-} \\ \textbf{(Devices in the)} \\ \textbf{Station}}} & MitM, Eavesdropping & Data Leakage & - Steal sensitive information from user & Encryption, Authentication, VPN & Low \\ 
\cline{2-6}
& Spoofing, Authentication system attack & Manipulation, System Corruption, Service Loss & \multirow{3}{*}{\makecell[{{p{7cm}}}]{- Bill another user for a ticket\\ - Prevent users from booking a ticket \\ - Prevent user reaching the legitimate pod}} & Encryption, Digital Signature & Low \\ 
\cline{2-6}
& Privacy violation & Data Leakage & \multirow{2}{*}{\makecell[{{p{7cm}}}]{- Tracking users pod’s accesses \\ - Profiling users}} & Data Anonymization & Low \\
\hline
\end{tabular}
}
\end{table*}

\section{Attacks Countermeasures}\label{sec:countermeasures}

The first line of protection of the system should follow the best security practices and security standards from the system's design phase. These standards are applied not only by industries but also by governments and researchers. Since Hyperloop is a novel technology, no security standards support its design phase. However, Hyperloop inherits different properties from other sectors, such as automotive, railway systems, and aviation. Therefore, the designer can partially rely on the existing standards, as highlighted in the Roadmap from the CEN-CENELEC~\cite{jtc20}. It includes general standards for cybersecurity, such as the \textit{ISO/IEC 27000} series~\cite{iso27000}, and more specific regulations such as \textit{IEC 62443}~\cite{IEC62443, leander2019applicability} regarding industrial automation and \textit{CLC/TS 50701}~\cite{CLC/TS50701} discussing cybersecurity of railways systems. In addition to them, we found useful other documents related to other fields. For instance, automotive security policies are collected in documents such as \textit{AUTOSAR}~\cite{furst2009autosar} and \textit{ISO/SAE 21434:2021}~\cite{iso21434}. Instead, security by design for industrial security is extensively discussed in the \textit{NCSC}~\cite{NCSC} secure design principles, which can represent an important read during the design of Hyperloop systems. %

Besides security standards, security mechanisms can be applied to increase the overall security level of the system and prevent attacks. Table~\ref{tab:secuirty} shows the list of the possible countermeasures related to each attack previously described.

To prevent MitM attacks, the most common solution is introducing encryption and authentication techniques in the communication channels to avoid data manipulation from unauthorized parties. Instead, DoS attacks are very challenging to prevent. Common strategies include resource managers correctly allocating the memory to the critical processes and redistributing the resources if a DoS attack is detected. Moreover, solutions employing cooperative control strategies may be adapted from related domains~\cite{zhao2021resilient}. To prevent network attacks, a common solution is the implementation of an Intrusion Detection System (IDS) to identify malicious behaviors early and make accurate decisions to protect the system~\cite{gao2019novel, xiao2021intrusion}. It is also possible to implement a firewall in strategic points of the network (e.g., the station or the different access points) to increase the perimetral protection of the network.

An important factor for preserving the system's function is to ensure the correct localization of every pod. For this reason, an attacker may try to spoof the pod's location to alternate and compromise Hyperloop. Other than the previous measures to avoid spoofings, like encryption and authentication, a common solution to prevent spoofing of communication is the introduction of a distance bounding protocol~\cite{rasmussen2010realization}. This protocol measures the physical distance of an entity by analyzing the expected response time. In this way, the receiver will detect the delay added by manipulating data or by a spoofed position. This protocol has been widely used to prevent relay attacks and GPS spoofing~\cite{wen2005countermeasures}.

Due to the system's distributed nature, the physical infrastructure of the Hyperloop also opens potential vulnerability surfaces. The most common solution to prevent physical tampering is integrating anti-tampering material equipped with sensors that identify attempts~\cite{desai2013interlocking}. Another solution is to employ anomaly detector software to identify the alteration of a process.

A standard solution to prevent user privacy leakages is to provide correct anonymization in all communications involving sensible data. Possible solutions include encryption and data aggregation. In the case of devices with limited capacity or time-constrained operations, differential privacy represents a viable solution~\cite{dwork2006differential}.

\section{Conclusions and Future Directions}\label{sec:conclusion}

Hyperloop represents an innovative and promising technology for mass transportation that is still under development.
Even if the Joint Technical Committee 20 is working on a standard for Hyperloop systems~\cite{jtc20}, currently, there are no official documents detailing the system's technical implementation, which does not allow precise analysis of the infrastructure. However, cybersecurity is a continuous analysis process, which must be seriously considered from the design phase. This requirement is emphasized since Hyperloop exhibits a large attack surface comprising network, communications, and physical processes.
Compromising the security of a transportation system can dramatically impact the safety of the passengers and the surrounding environment.

This document represents the first work in the direction of the cybersecurity analysis of the Hyperloop system. Specifically, we identified the communication characterizing the Hyperloop system and investigated possible threats on the different parts of the infrastructure. Then, we discussed the best practices, countermeasures, and standards to prevent potential catastrophic attacks. %
Furthermore, we discussed the privacy issues related to users and the infrastructure. For each of the identified vulnerabilities, we also proposed possible countermeasures.

This analysis should be deepened in future works using more precise Hyperloop technical details. In addition, we expect that academia and industries will develop more reliable communication and efficient protocols to support the Hyperloop high-speed communication shortly. Future security researchers must also consider and include such novel technologies.

\section*{Acknowledgement}

Denis Donadel is supported by Omitech S.r.l, that we want to thank. We also thank Elisa Turrin (aka \textit{Upata}) for the graphical illustration.

\bibliographystyle{IEEEtran}
\bibliography{biblio}

\begin{thebibliography}{10}
\providecommand{\url}[1]{#1}
\csname url@samestyle\endcsname
\providecommand{\newblock}{\relax}
\providecommand{\bibinfo}[2]{#2}
\providecommand{\BIBentrySTDinterwordspacing}{\spaceskip=0pt\relax}
\providecommand{\BIBentryALTinterwordstretchfactor}{4}
\providecommand{\BIBentryALTinterwordspacing}{\spaceskip=\fontdimen2\font plus
\BIBentryALTinterwordstretchfactor\fontdimen3\font minus
  \fontdimen4\font\relax}
\providecommand{\BIBforeignlanguage}[2]{{%
\expandafter\ifx\csname l@#1\endcsname\relax
\typeout{** WARNING: IEEEtran.bst: No hyphenation pattern has been}%
\typeout{** loaded for the language `#1'. Using the pattern for}%
\typeout{** the default language instead.}%
\else
\language=\csname l@#1\endcsname
\fi
#2}}
\providecommand{\BIBdecl}{\relax}
\BIBdecl

\bibitem{musk2013hyperloop}
E.~Musk, ``Hyperloop alpha,'' \emph{SpaceX: Hawthorne, CA, USA}, 2013.

\bibitem{polunsky2017homeland}
S.~M. Polunsky, ``Homeland security and texas’ high-speed rail,''
  \emph{Journal of Transportation Security}, vol.~10, no.~3, pp. 73--86, 2017.

\bibitem{janic2021estimation}
M.~Jani{\'c}, ``Estimation of direct energy consumption and co2 emission by
  high speed rail, transrapid maglev and hyperloop passenger transport
  systems,'' \emph{International Journal of Sustainable Transportation},
  vol.~15, no.~9, pp. 696--717, 2021.

\bibitem{MTST81420}
R.~Özbek and M.~Y. Çodur, ``Comparison of hyperloop and existing transport
  vehicles in terms of security and costs,'' \emph{Modern Transportation
  Systems and Technologies}, vol.~7, no.~3, pp. 5--29, 2021.

\bibitem{hyperloopWikipedia}
\BIBentryALTinterwordspacing
Hyperloop companies. [Accessed on June 2023]. [Online]. Available:
  \url{https://en.wikipedia.org/wiki/Hyperloop#Hyperloop_companies}
\BIBentrySTDinterwordspacing

\bibitem{hyperloopitalia}
\BIBentryALTinterwordspacing
Hyperloopitalia. [Accessed on June 2023]. [Online]. Available:
  \url{https://hyperloopitalia.com/}
\BIBentrySTDinterwordspacing

\bibitem{delft}
\BIBentryALTinterwordspacing
Delft hyperloop. [Accessed on June 2023]. [Online]. Available:
  \url{https://www.delfthyperloop.nl/}
\BIBentrySTDinterwordspacing

\bibitem{htt}
\BIBentryALTinterwordspacing
Hyperloop transportation technology. [Accessed on June 2023]. [Online].
  Available: \url{https://www.hyperlooptt.com/}
\BIBentrySTDinterwordspacing

\bibitem{duo2022survey}
W.~Duo, M.~Zhou, and A.~Abusorrah, ``A survey of cyber attacks on cyber
  physical systems: Recent advances and challenges,'' \emph{IEEE/CAA Journal of
  Automatica Sinica}, vol.~9, no.~5, pp. 784--800, 2022.

\bibitem{thakur2016impact}
K.~Thakur, M.~L. Ali, N.~Jiang, and M.~Qiu, ``Impact of cyber-attacks on
  critical infrastructure,'' in \emph{2016 IEEE 2nd International Conference on
  Big Data Security on Cloud (BigDataSecurity), IEEE International Conference
  on High Performance and Smart Computing (HPSC), and IEEE International
  Conference on Intelligent Data and Security (IDS)}.\hskip 1em plus 0.5em
  minus 0.4em\relax IEEE, 2016, pp. 183--186.

\bibitem{falliere2011w32}
N.~Falliere, L.~O. Murchu, and E.~Chien, ``W32. stuxnet dossier,'' \emph{White
  paper, symantec corp., security response}, vol.~5, no.~6, p.~29, 2011.

\bibitem{reutershacker}
\BIBentryALTinterwordspacing
Reuters. Hackers breach iran rail network, disrupt service. [Accessed on June
  2023]. [Online]. Available:
  \url{https://www.reuters.com/world/middle-east/hackers-breach-iran-rail-network-disrupt-service-2021-07-09/}
\BIBentrySTDinterwordspacing

\bibitem{noland2021prospects}
J.~K. N{\o}land, ``Prospects and challenges of the hyperloop transportation
  system: A systematic technology review,'' \emph{IEEE Access}, vol.~9, pp.
  28\,439--28\,458, 2021.

\bibitem{chafii2023twelve}
M.~Chafii, L.~Bariah, S.~Muhaidat, and M.~Debbah, ``Twelve scientific
  challenges for 6g: Rethinking the foundations of communications theory,''
  \emph{IEEE Communications Surveys \& Tutorials}, 2023.

\bibitem{gkoumas2021hyperloop}
K.~Gkoumas, ``Hyperloop academic research: A systematic review and a taxonomy
  of issues,'' \emph{applied sciences}, vol.~11, no.~13, p. 5951, 2021.

\bibitem{brusyanin2014basic}
D.~Brusyanin and S.~Vikharev, ``The basic approach in designing of the
  functional safety index for transport infrastructure,'' \emph{Contemporary
  Engineering Sciences}, vol.~7, no.~6, pp. 287--292, 2014.

\bibitem{virgin}
\BIBentryALTinterwordspacing
Virgin hyperloop. [Accessed on June 2023]. [Online]. Available:
  \url{https://web.archive.org/web/20220326090510/https://virginhyperloop.com/}
\BIBentrySTDinterwordspacing

\bibitem{korea}
\BIBentryALTinterwordspacing
Hyperloop achieves 1000 km/h speed in korea, days after virgin passenger test.
  independent 2020. [Accessed on March 2024]. [Online]. Available:
  \url{https://www.independent.co.uk/tech/hyperloop-korea-speed-record-korail-virgin-b1721942.html}
\BIBentrySTDinterwordspacing

\bibitem{spain}
\BIBentryALTinterwordspacing
Spain’s zeleros raises 7m€ in financing to lead the development of
  hyperloop in europe. [Accessed on March 2024]. [Online]. Available:
  \url{https://zeleros.com/}
\BIBentrySTDinterwordspacing

\bibitem{poland}
\BIBentryALTinterwordspacing
Hyper poland reveals its magrail transport technology. [Accessed on March
  2024]. [Online]. Available:
  \url{https://ecotechdaily.net/hyper-poland-reveals-their-magrail-transport-technology/}
\BIBentrySTDinterwordspacing

\bibitem{france}
\BIBentryALTinterwordspacing
Test half-scale test track in droux. [Accessed on March 2024]. [Online].
  Available:
  \url{https://web.archive.org/web/20220303184714/https://www.transpod.com/test-facility/}
\BIBentrySTDinterwordspacing

\bibitem{swiss}
\BIBentryALTinterwordspacing
Building the next mode of transportation: Hyperloop. [Accessed on March 2024].
  [Online]. Available: \url{https://swisspod.ch/}
\BIBentrySTDinterwordspacing

\bibitem{gkoumas2021evidence}
K.~Gkoumas and M.~Christou, ``Evidence-based challenges for hyperloop
  deployment: A taxonomy of research issues based on bibliographic research,''
  \emph{Transportation Research Board 100th Annual Meeting}, 2021.

\bibitem{opgenoord2018aerodynamic}
M.~M. Opgenoord and P.~C. Caplan, ``Aerodynamic design of the hyperloop
  concept,'' \emph{Aiaa Journal}, vol.~56, no.~11, pp. 4261--4270, 2018.

\bibitem{nick2020computational}
N.~Nick and Y.~Sato, ``Computational fluid dynamics simulation of hyperloop pod
  predicting laminar--turbulent transition,'' \emph{Railway Engineering
  Science}, vol.~28, pp. 97--111, 2020.

\bibitem{nowacki2019assessment}
M.~Nowacki, D.~Olejniczak, and J.~Markowski, ``Assessment of medium parameters
  in a closed space for a hyperloop transport capsule with reference to
  reducing the energy demand of a transport system,'' in \emph{E3S Web of
  Conferences}, vol. 108.\hskip 1em plus 0.5em minus 0.4em\relax EDP Sciences,
  2019, p. 01032.

\bibitem{seo2020study}
K.-Y. Seo, C.-B. Park, G.~Jeong, J.-B. Lee, T.~Kim, and H.-W. Lee, ``A study on
  the design of propulsion/levitation/guidance integrated dslim with
  non-symmetric structure,'' \emph{AIP Advances}, vol.~10, no.~2, 2020.

\bibitem{chaidez2019levitation}
E.~Chaidez, S.~P. Bhattacharyya, and A.~N. Karpetis, ``Levitation methods for
  use in the hyperloop high-speed transportation system,'' \emph{Energies},
  vol.~12, no.~21, p. 4190, 2019.

\bibitem{heat}
F.~Lluesma-Rodr{\'\i}guez, F.~Alcantara-Avila, M.~J. P{\'e}rez-Quiles, and
  S.~Hoyas, ``A code for simulating heat transfer in turbulent channel flow,''
  \emph{Mathematics}, vol.~9, no.~7, p. 756, 2021.

\bibitem{riviera2018high}
M.~Riviera, ``High-speed trains comparison to hyperloop: energy, sustainability
  and safety analysis hyperloop integrations to reach the noah concept,'' Ph.D.
  dissertation, Politecnico di Torino, 2018.

\bibitem{van2018analysis}
K.~van Goeverden, D.~Milakis, M.~Janic, and R.~Konings, ``Analysis and
  modelling of performances of the hl (hyperloop) transport system,''
  \emph{European Transport Research Review}, vol.~10, no.~2, pp. 1--17, 2018.

\bibitem{mateu2021setting}
J.~M. Mateu, P.~M. Fern{\'a}ndez, and R.~I. Franco, ``Setting safety
  foundations in the hyperloop: A first approach to preliminary hazard analysis
  and safety assurance system,'' \emph{Safety science}, vol. 142, p. 105366,
  2021.

\bibitem{han2020wireless}
L.~Han, H.~Wu, and X.~Chen, ``Wireless network architecture for evacuated tube
  transportation system,'' \emph{China Communications}, vol.~17, no.~10, pp.
  206--217, 2020.

\bibitem{qiu2020broadband}
C.~Qiu, L.~Liu, B.~Han, J.~Zhang, Z.~Li, and T.~Zhou, ``Broadband wireless
  communication systems for vacuum tube high-speed flying train,''
  \emph{Applied Sciences}, vol.~10, no.~4, p. 1379, 2020.

\bibitem{zhang2020concepts}
J.~Zhang, L.~Liu, B.~Han, Z.~Li, T.~Zhou, K.~Wang, D.~Wang, and B.~Ai,
  ``Concepts on train-to-ground wireless communication system for hyperloop:
  Channel, network architecture, and resource management,'' \emph{Energies},
  vol.~13, no.~17, p. 4309, 2020.

\bibitem{tavsanoglu2021concepts}
A.~Tavsanoglu, C.~Briso, D.~Carmena-Cabanillas, and R.~B. Arancibia, ``Concepts
  of hyperloop wireless communication at 1200 km/h: 5g, wi-fi, propagation,
  doppler and handover,'' \emph{Energies}, vol.~14, no.~4, p. 983, 2021.

\bibitem{huang2021optical}
X.~Huang, F.~Yang, J.~Song, and Z.~Han, ``An optical communication approach for
  ultra-high-speed train running in evacuated tube: Potentials and
  challenges,'' \emph{IEEE Wireless Communications}, vol.~28, no.~3, pp.
  70--76, 2021.

\bibitem{hedhly2023intelligent}
W.~Hedhly, O.~Amin, M.-S. Alouini, and B.~Shihada, ``Intelligent reflecting
  surfaces assisted hyperloop wireless communication network,'' \emph{IEEE
  Transactions on Mobile Computing}, 2023.

\bibitem{li2024scheduling}
P.~Li, Y.~Niu, H.~Wu, Z.~Han, Y.~Wang, N.~Wang, Z.~Zhong, and B.~Ai,
  ``Scheduling of millimeter wave communications for ultra-high-speed vacuum
  tube train,'' \emph{IEEE Transactions on Vehicular Technology}, 2024.

\bibitem{hedhly2021hyperloop}
W.~Hedhly, O.~Amin, B.~Shihada, and M.-S. Alouini, ``Hyperloop communications:
  Challenges, advances, and approaches,'' \emph{IEEE Open Journal of the
  Communications Society}, vol.~2, pp. 2413--2435, 2021.

\bibitem{dudnikov2017advantages}
E.~Dudnikov, ``Advantages of a new hyperloop transport technology,'' in
  \emph{2017 Tenth International Conference Management of Large-Scale System
  Development (MLSD)}.\hskip 1em plus 0.5em minus 0.4em\relax IEEE, 2017, pp.
  1--4.

\bibitem{chin2015open}
J.~C. Chin and J.~S. Gray, ``Open-source conceptual sizing models for the
  hyperloop passenger pod,'' in \emph{56th AIAA/ASCE/AHS/ASC Structures,
  Structural Dynamics, and Materials Conference}, 2015, p. 1587.

\bibitem{ghosal2021truck}
A.~Ghosal, S.~U. Sagong, S.~Halder, K.~Sahabandu, M.~Conti, R.~Poovendran, and
  L.~Bushnell, ``Truck platoon security: State-of-the-art and road ahead,''
  \emph{Computer Networks}, vol. 185, p. 107658, 2021.

\bibitem{singh2021board}
N.~Singh, J.~Karhade, I.~Bhattacharya, P.~Saraf, P.~Kattamuri, and A.~M.
  Parimi, ``On-board electrical, electronics and pose estimation system for
  hyperloop pod design,'' in \emph{2021 7th International Conference on
  Control, Automation and Robotics (ICCAR)}.\hskip 1em plus 0.5em minus
  0.4em\relax IEEE, 2021, pp. 223--230.

\bibitem{jtc20}
``{JTC 20 - CEN/CLC/TR 17912:2023 --- Hyperloop systems - Standards Inventory
  and Roadmap},'' European Committee For Electrotechnical Standardization,
  Brussels, Belgium, Standard, Jan. 2023.

\bibitem{iso27000}
``{ISO/IEC 27000:2018 --- Information technology --- Security techniques},''
  International Organization for Standardization, Geneva, CH, Standard, Feb.
  2018.

\bibitem{IEC62443}
``{ISA/IEC 62443 --- Industrial communication networks --- Network and system
  security},'' International Electrotechnical Commission, Geneva, CH, Standard,
  Feb. 2019.

\bibitem{leander2019applicability}
B.~Leander, A.~{\v{C}}au{\v{s}}evi{\'c}, and H.~Hansson, ``Applicability of the
  iec 62443 standard in industry 4.0/iiot,'' in \emph{Proceedings of the 14th
  International Conference on Availability, Reliability and Security}, 2019,
  pp. 1--8.

\bibitem{CLC/TS50701}
``{CLC/TS50701 --- Railway applications --- Cybersecurity},'' European
  Committee For Electrotechnical Standardization, Brussels, Belgium, Standard,
  Aug. 2023.

\bibitem{furst2009autosar}
S.~F{\"u}rst, J.~M{\"o}ssinger, S.~Bunzel, T.~Weber, F.~Kirschke-Biller,
  P.~Heitk{\"a}mper, G.~Kinkelin, K.~Nishikawa, and K.~Lange, ``Autosar--a
  worldwide standard is on the road,'' in \emph{14th International VDI Congress
  Electronic Systems for Vehicles, Baden-Baden}, vol.~62, no.~5.\hskip 1em plus
  0.5em minus 0.4em\relax Citeseer, 2009.

\bibitem{iso21434}
``{ISO/SAE 21434:2021 --- Road vehicles --- Cybersecurity engineering},''
  International Organization for Standardization, Geneva, CH, Standard, Aug.
  2021.

\bibitem{NCSC}
``{NCSC --- Secure design principles},'' National Cyber Security Center,
  London, UK, Standard, May 2019.

\bibitem{zhao2021resilient}
H.~Zhao, X.~Dai, L.~Ding, D.~Cui, J.~Ding, and T.~Chai, ``Resilient cooperative
  control for high-speed trains under denial-of-service attacks,'' \emph{IEEE
  Transactions on Vehicular Technology}, vol.~70, no.~12, pp. 12\,427--12\,436,
  2021.

\bibitem{gao2019novel}
B.~Gao and B.~Bu, ``A novel intrusion detection method in train-ground
  communication system,'' \emph{IEEE Access}, vol.~7, pp. 178\,726--178\,743,
  2019.

\bibitem{xiao2021intrusion}
X.~Xiao, X.~Ma, Y.~Hui, Z.~Yin, T.~H. Luan, and Y.~Wu, ``Intrusion detection
  for high-speed railway system: a faster r-cnn approach,'' in \emph{2021 IEEE
  94th Vehicular Technology Conference (VTC2021-Fall)}.\hskip 1em plus 0.5em
  minus 0.4em\relax IEEE, 2021, pp. 1--5.

\bibitem{rasmussen2010realization}
K.~B. Rasmussen and S.~Capkun, ``Realization of $\{$RF$\}$ distance bounding,''
  in \emph{19th USENIX Security Symposium (USENIX Security 10)}, 2010.

\bibitem{wen2005countermeasures}
H.~Wen, P.~Y.-R. Huang, J.~Dyer, A.~Archinal, and J.~Fagan, ``Countermeasures
  for gps signal spoofing,'' in \emph{Proceedings of the 18th international
  technical meeting of the satellite division of the institute of navigation
  (ION GNSS 2005)}, 2005, pp. 1285--1290.

\bibitem{desai2013interlocking}
A.~R. Desai, M.~S. Hsiao, C.~Wang, L.~Nazhandali, and S.~Hall, ``Interlocking
  obfuscation for anti-tamper hardware,'' in \emph{Proceedings of the eighth
  annual cyber security and information intelligence research workshop}, 2013,
  pp. 1--4.

\bibitem{dwork2006differential}
C.~Dwork, ``Differential privacy,'' in \emph{International colloquium on
  automata, languages, and programming}.\hskip 1em plus 0.5em minus 0.4em\relax
  Springer, 2006, pp. 1--12.

\end{thebibliography}

\end{document}